\documentclass[final]{aipproc}

\layoutstyle{8x11single}
\usepackage{graphicx}
\usepackage{subfig}
\usepackage{array}

\begin{document}

\title{A comparison of Monte Carlo generators}



\author{Tomasz Golan}{address = {Institute for Theoretical Physics, Wroc{\l}aw University, Plac Maxa Borna 9, 50-204 Wroc{\l}aw, Poland}}

\begin{abstract}

A comparison of GENIE, NEUT, NUANCE, and NuWro Monte Carlo neutrino event generators is presented using a set of four observables: protons multiplicity, total visible energy, most energetic proton momentum, and $\pi^+$ two-dimensional energy vs cosine distribution.

\end{abstract}

\maketitle

\section{Introduction}

Monte Carlo (MC) generators are important tools in the investigation of neutrino measurements \cite{Gallagher:2009zza}. Regardless a measurement method used in an experiment, the particles seen in the detector are the ones produced in a neutrino-nucleus scattering, usually affected by the final state interactions (FSI). To interpret the observation one needs to use MC simulations.

There are several MC neutrino event generators and they have many common features. Each distinguishes quasi-elastic scattering, pion production mainly through $\Delta$ resonance excitation (the isobar model for single pion production with a non-resonant background \cite{Juszczak:2005zs} in NuWro or Rein-Sehgal model \cite{Rein:1981} in other generators), more inelastic processes (in neutrino MC community usually called deep inelastic scattering), and coherent pion production for both neutral current (NC) and charged current (CC) interactions. For all dynamics, but the coherent one, the impulse approximation is assumed, so the scattering occurs on an individual nucleon from the nucleus. Particles created in the primary interaction are propagated through the nucleus using the intra-nuclear cascade (INC) model \cite{Metropolis:1958}.

The basic idea behind the MC codes is similar for all event generators, however, the results can differ significantly on the level of both the primary vertex and the final state interactions \cite{Antonello:2009ca}. This gives a motivation to investigate more carefully and improve the performance of MC generators. However, we should keep in mind that neutrino-nucleus cross sections are known with a precision of $20-30\%$, and MC cannot do better.

Because there is limited set of data available to test the codes, comparisons between MC generators are very useful. Such approach has an extra advantage of a freedom in defining the observables, which may be not easily measurable, but still useful to crosscheck the generators. In this paper the results obtain using GENIE \cite{Andreopoulos:2009rq}, NUANCE \cite{Casper:2002sd}, NEUT \cite{Hayato:2009zz}, and NuWro \cite{Golan:2012wx} are presented. The first of them is currently being used in experiments, such as T2K, NOvA, MINOS and MINERvA. NEUT is an official MC of the T2K collaboration, and NUANCE in the MiniBooNE experiment.

\section{Comparisons}

\subsection{A protons multiplicity}

\begin{figure}

\captionsetup{type=figure}
\begin{tabular}{cc}
 \subfloat[]{\includegraphics[width = 0.5\columnwidth]{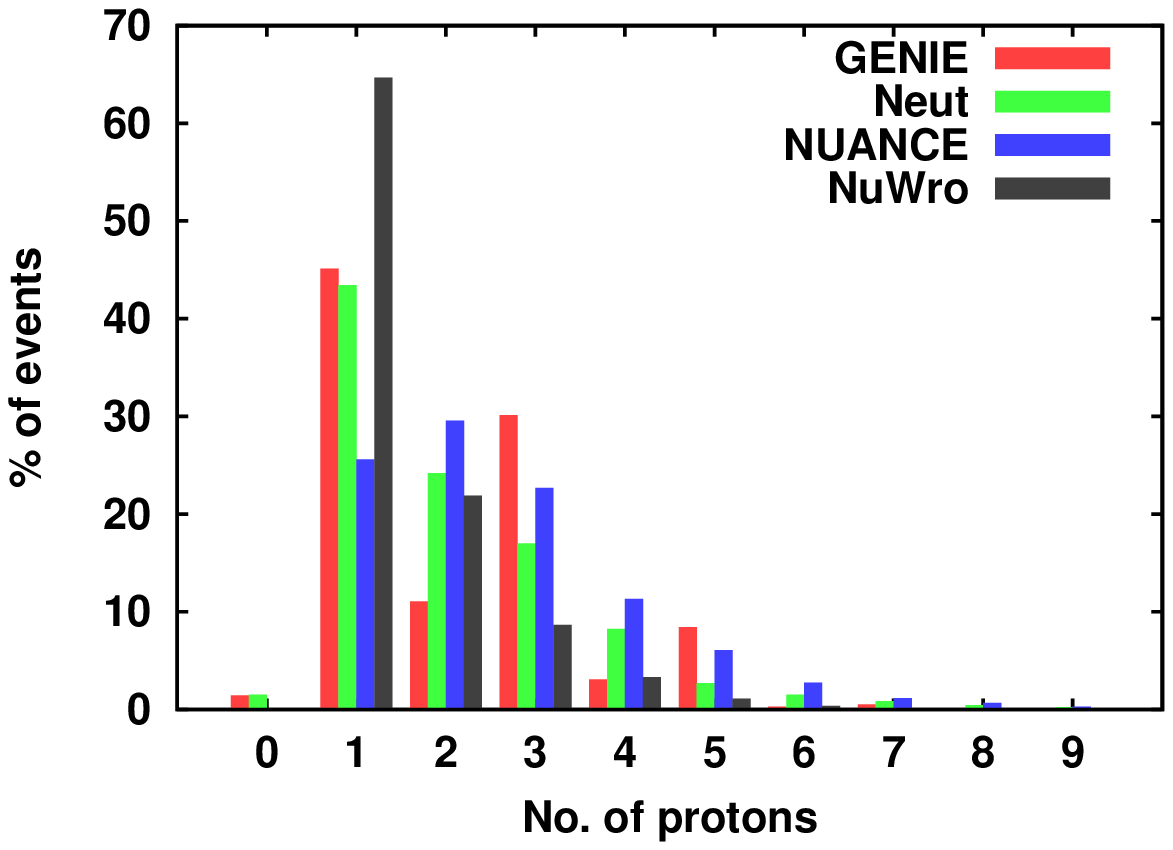}} &
 \subfloat[]{\includegraphics[width = 0.5\columnwidth]{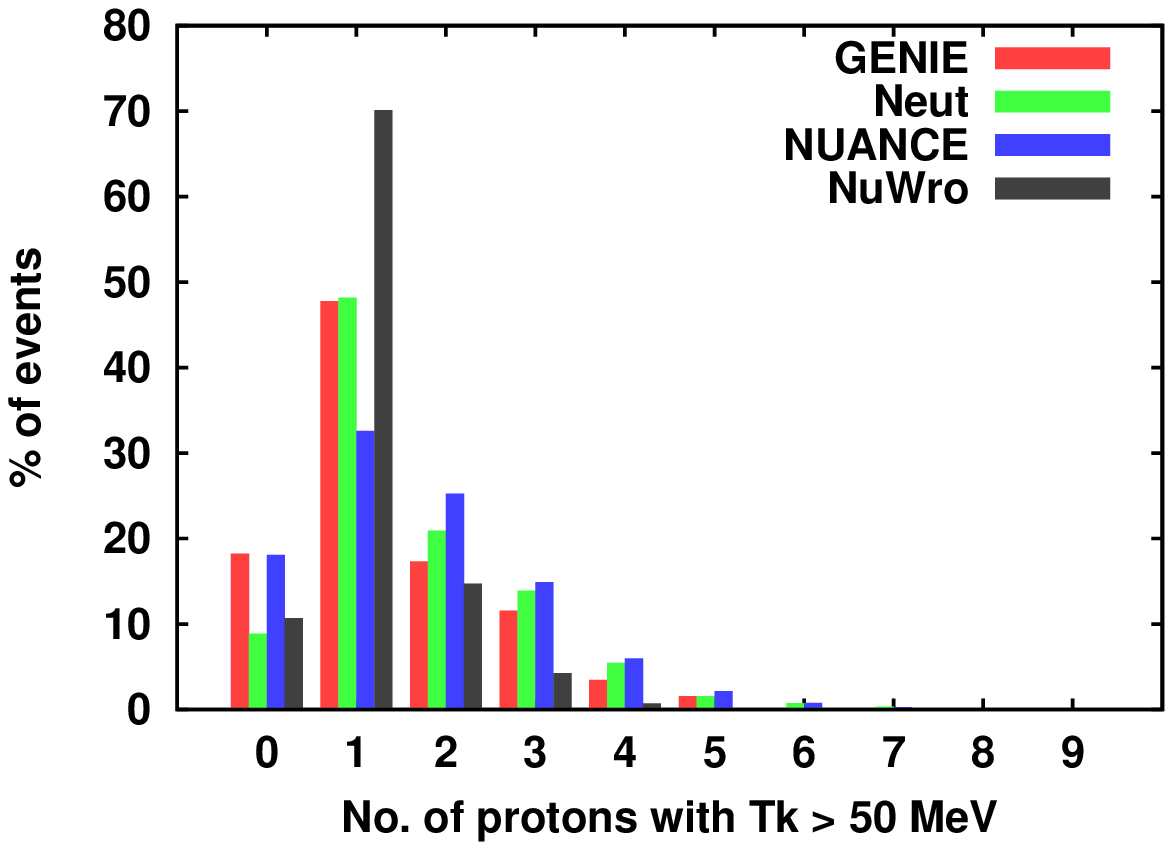}} \\
 \subfloat[]{\includegraphics[width = 0.5\columnwidth]{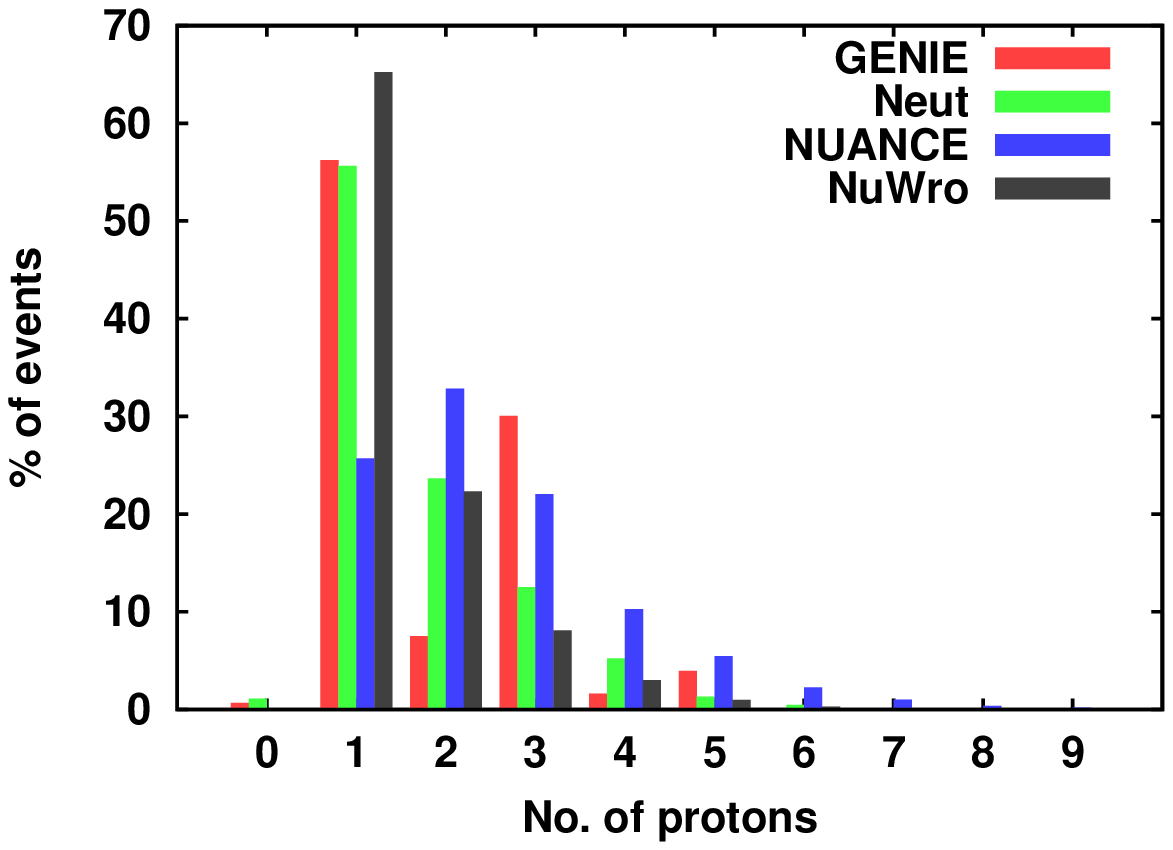}} &
 \subfloat[]{\includegraphics[width = 0.5\columnwidth]{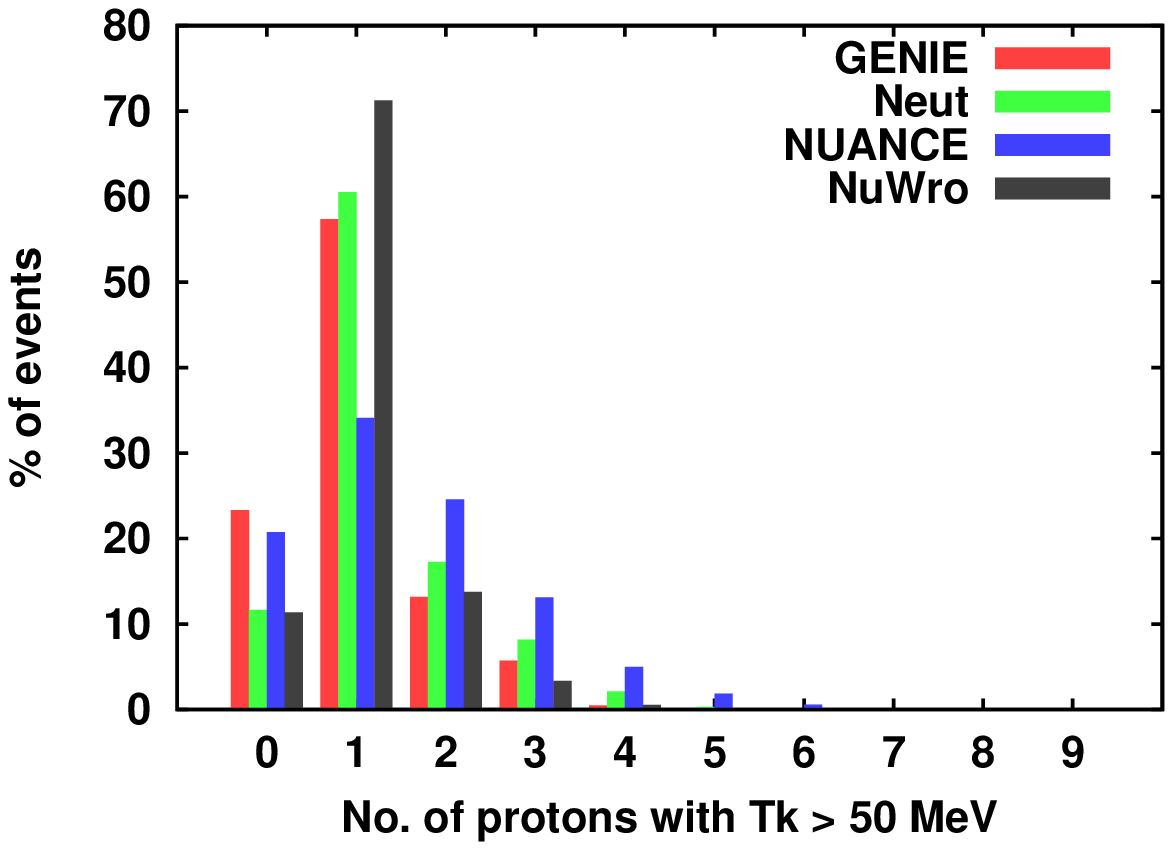}} \\
\end{tabular} 

\label{fig: DE11}
\caption{Protons multiplicity in the neutrino-argon CC interaction with no meson in a final state for the neutrino energy: (a), (b) $E_\nu = 3$~GeV and (c), (d) $E_\nu = 1$~GeV, for the cases (a), (c) with no cut and (b),(d) with a cut on the proton kinetic energy $T_k > 50$~MeV.}

\end{figure}

The first investigated observable is protons multiplicity in the neutrino-argon CC interaction for two values of the incident neutrino energy: $E_\nu = 1$~GeV and $E_\nu = 3$~GeV (see Fig.~\ref{fig: DE11}). Only events with no meson in the final state are taken into account. The results are obtain in two cases: either with a cut on the proton kinetic energy ($T_k > 50$~MeV) or without any cut. The predictions depend mainly on the nucleons cascade, however, a significant impact comes also from a model of pion absorption, giving a significant contribution to multi-nucleon events.

The differences between MC predictions are quite large. In the case of $3$~GeV energy NuWro produces more events with only one proton in the final state than other generators. Probably the reason of that lies in a treatment of the formation zone (FZ) effect \cite{Golan:2012wx}, which decreases a probability of re-interactions. The FZ effect is smaller for lower energies, so the difference at $E_\nu = 1$~GeV is also smaller. GENIE and NEUT have a quite good agreement with each other when the cut on the proton kinetic energy is applied. In all the cases NUANCE produces more multi-proton events than other generators.

Note that when the preliminary results for protons multiplicity from the ArgoNeuT experiment \cite{Partyka:NUINT12} become available, it will be possible to confront the MC predictions with them.

\subsection{Total visible energy}

\begin{figure}
\includegraphics[width = 0.5\columnwidth]{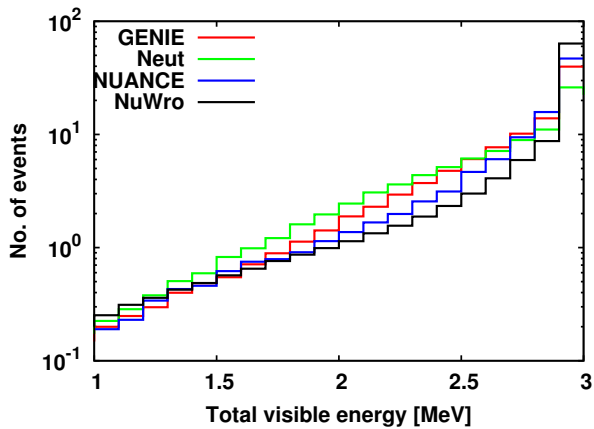}
\label{fig: DE31}
\caption{Total visible energy (the sum of kinetic energies of protons and total energies of mesons and charged leptons) in the neutrino-argon CC interaction for the neutrino energy $E_\nu = 3$~GeV.}
\end{figure}

Total visible energy, defined as the sum of the kinetic energies of all the protons and total energies of all the mesons and charged lepton, for neutrino energy $E_\nu = 3$~GeV, in the case of neutrino-argon CC scattering is presented in Fig.~\ref{fig: DE31}. The loss of the initial energy is related to the number of neutrons in the final state and the treatment of losing energy by the nucleons due to the FSI effects. The predictions, again, are not satisfactorily consistent.

As it was mentioned before, NuWro uses a model of FZ, which highly depends on the incident neutrino energy. For $E_\nu = 3$~GeV it significantly decreases a number of re-interactions and it results lower energy loss, comparing to other generators. The highest energy loss is predicted by NEUT, and two other generators predictions are in the middle.

Having in mind the protons multiplicity disagreement, one can assume the same situation to hold true also for neutrons. It is probably the main reason of the differences in the MC results. However, it would be also interesting to investigate the total energy carrying by all the particles in a final state to check the size of the difference in modeling nuclear potential in MC generators.

\subsection{Most energetic proton momentum}

\begin{figure}

\captionsetup{type=figure}
\begin{tabular}{cc}
 \subfloat[]{\includegraphics[width = 0.5\columnwidth]{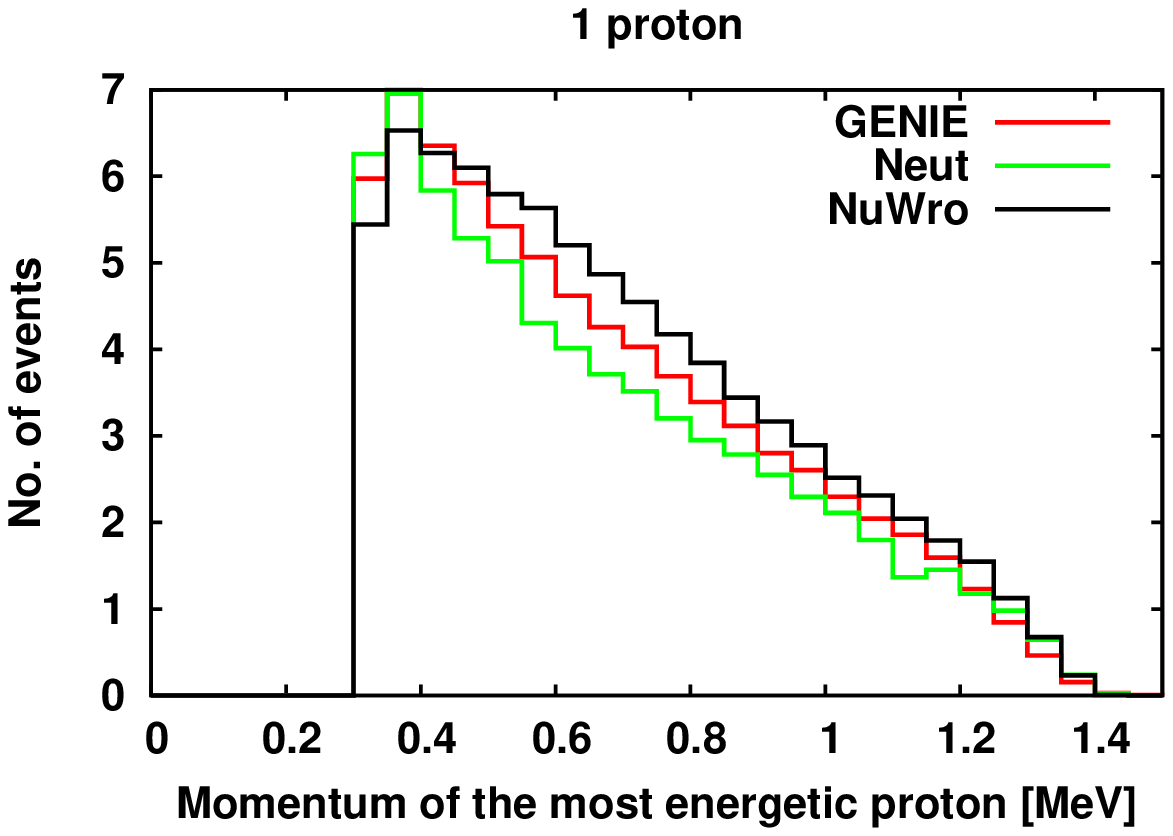}} &
 \subfloat[]{\includegraphics[width = 0.5\columnwidth]{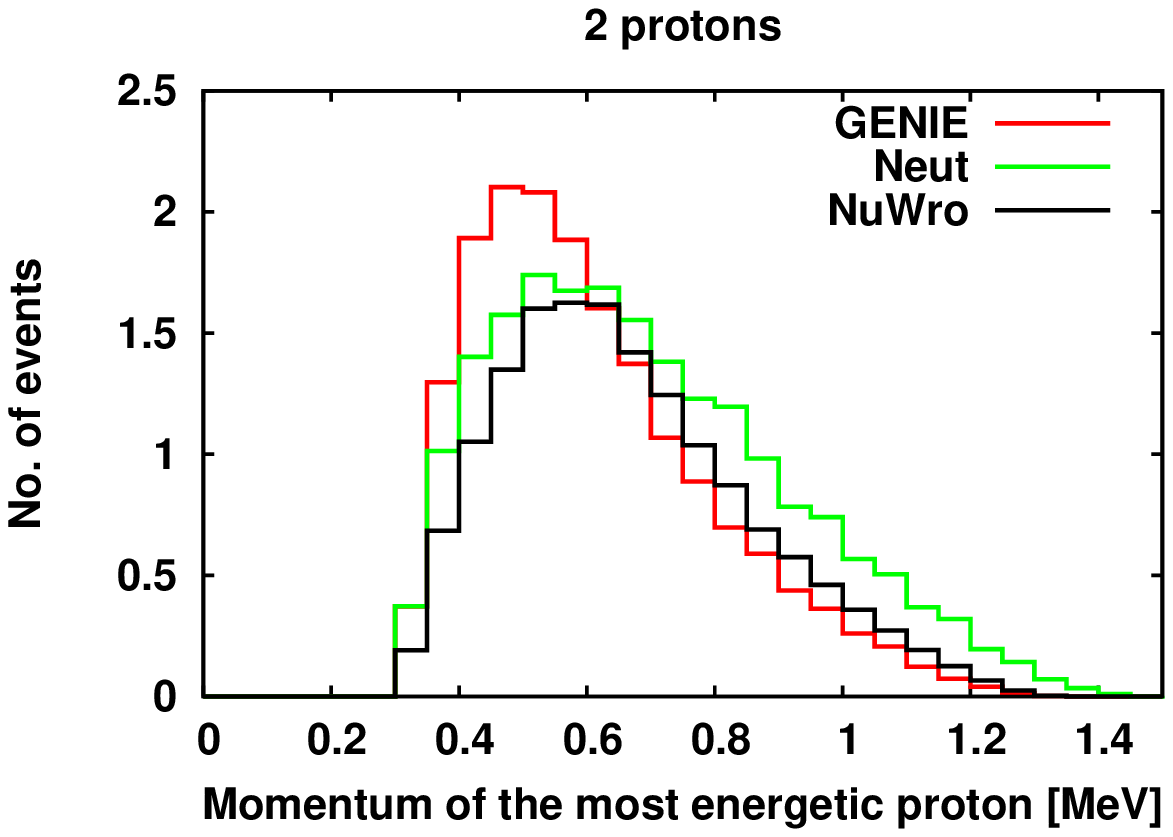}} \\
 \subfloat[]{\includegraphics[width = 0.5\columnwidth]{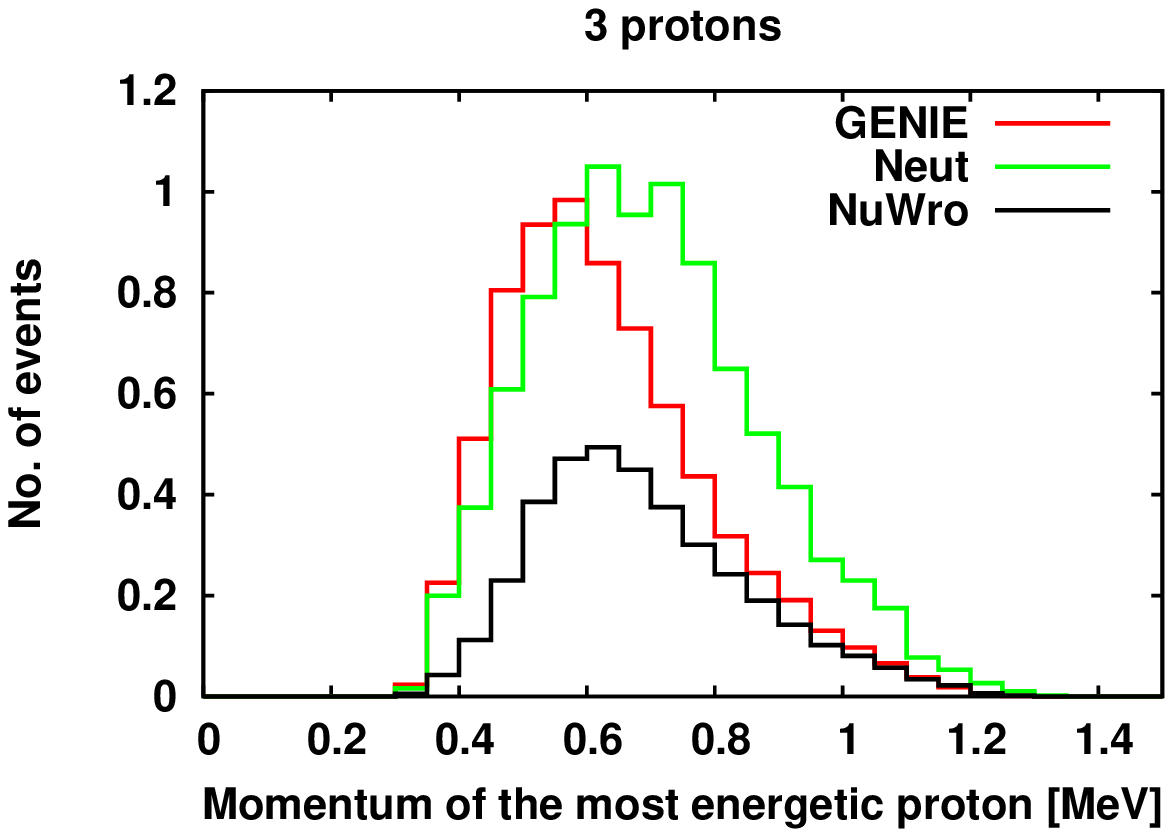}} &
 \subfloat[]{\includegraphics[width = 0.5\columnwidth]{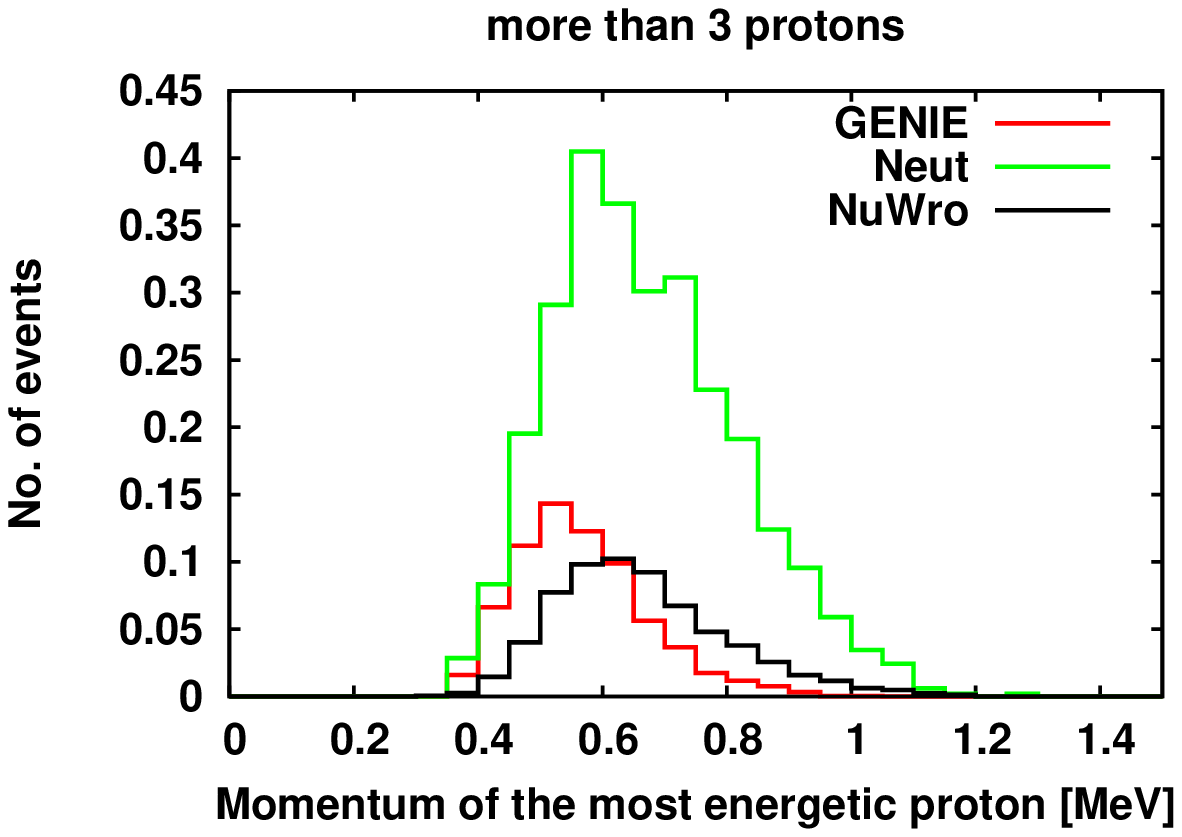}} \\
\end{tabular}

\label{fig: E13}
\caption{Most energetic proton momentum in the neutrino-argon CC interaction with no meson in a final state for the neutrino energy 1~GeV for events with: (a) one, (b) two, (c) three, and (d) more protons in the final state, normalized to 100 events, with a cut on the proton kinetic energies $T_k > 50$~MeV.}

\end{figure}

The next presented observable is the momentum of the most energetic proton in the neutrino-argon CC interaction for the neutrino energy $E_\nu = 1$~GeV. Only the events with no meson in the final state are investigated. In Fig.~\ref{fig: E13} the MC predictions are shown in four cases: events with only one, two, three or more protons in the final state. Together all they are normalized to 100 events. A cut on the proton kinetic energy $T_k > 50$~MeV was applied. In opposite to protons multiplicity, this results depends not only on the cross sections model used in nucleons cascade, but the kinematics is also important.

There is quite good agreement for all MC generators in the case of events with only one proton in the final state. When the protons number becomes larger, the differences in predictions start to increase, both in the shape and in the normalization. The crucial impact on the distributions for multi-proton events has a kinematics of the elastic nucleon-nucleon scattering and the pion absorption process. The treatment of the nuclear potential influence on the nucleons energy may also be important.

\subsection{$\pi^+$ two-dimensional energy vs cosine distribution}

\begin{figure}
\centering
\captionsetup{type=figure}
\begin{tabular}{cc}
 \subfloat[]{\includegraphics[width = 0.5\columnwidth]{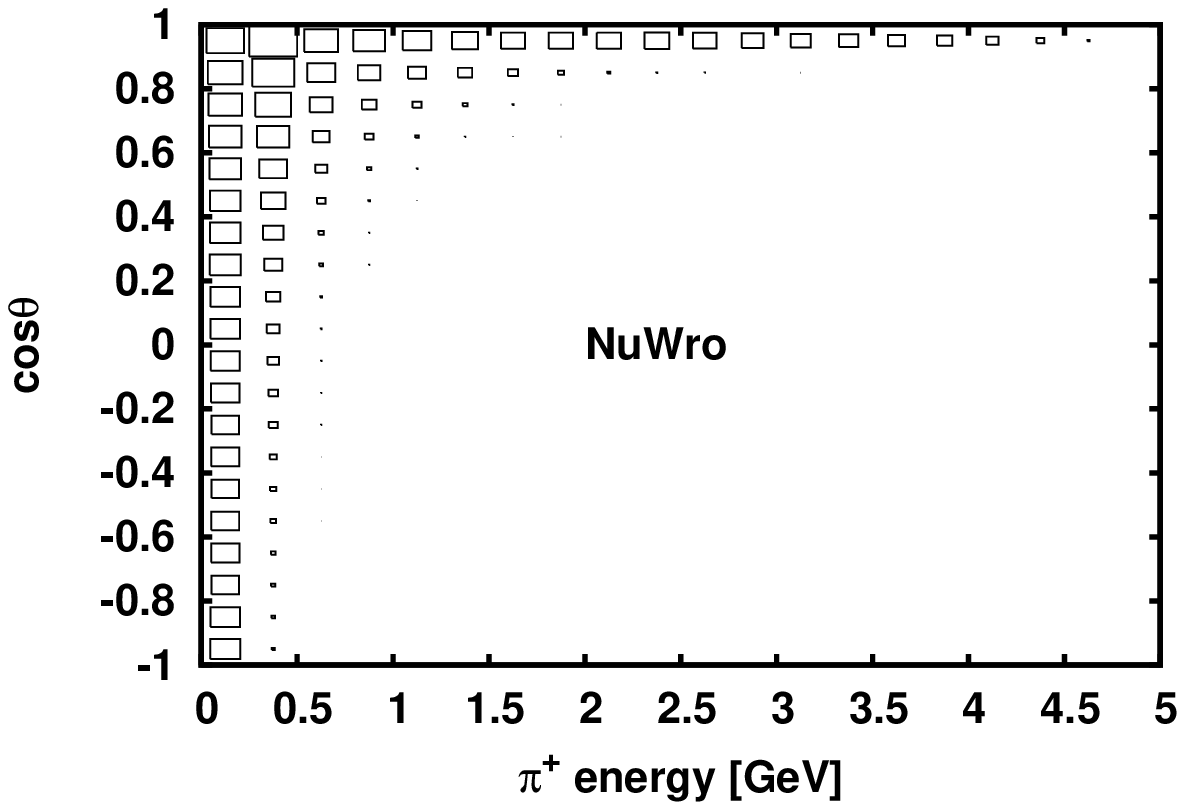}} &
 \subfloat[]{\includegraphics[width = 0.5\columnwidth]{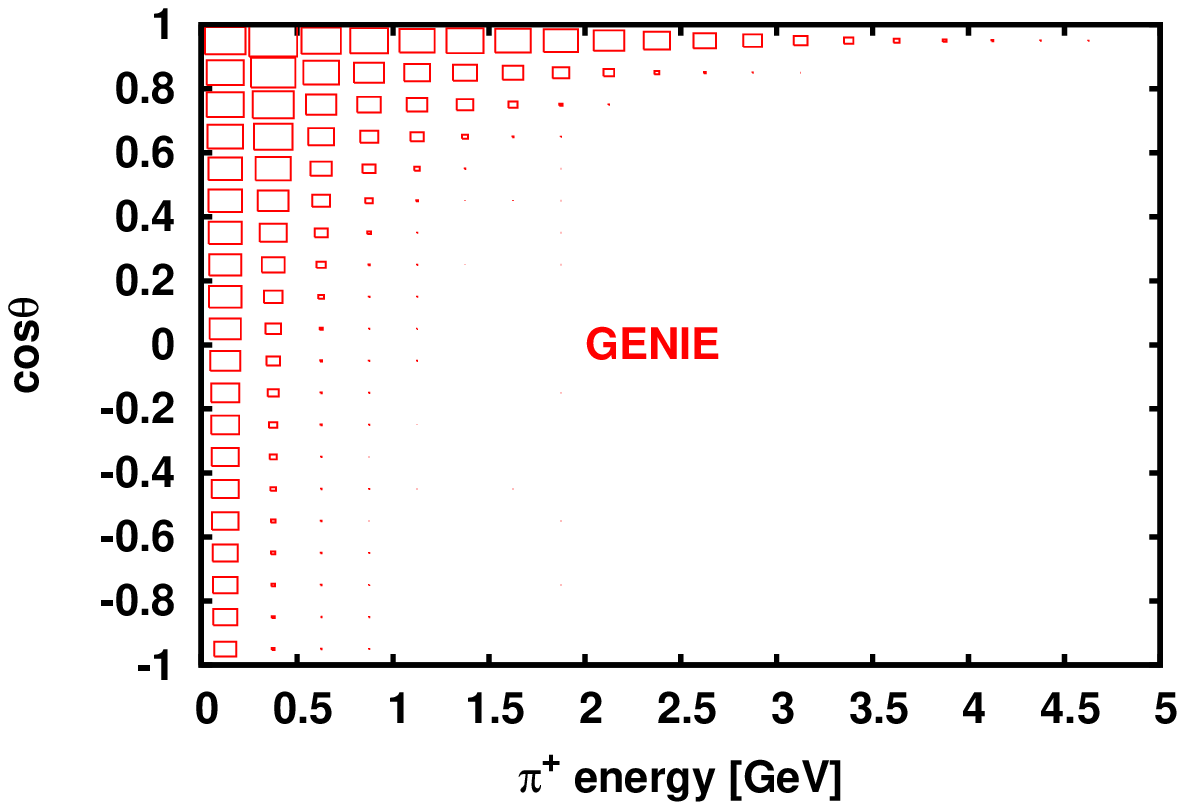}} \\
 \multicolumn{2}{c}{\subfloat[]{\centering\includegraphics[width = 0.5\columnwidth]{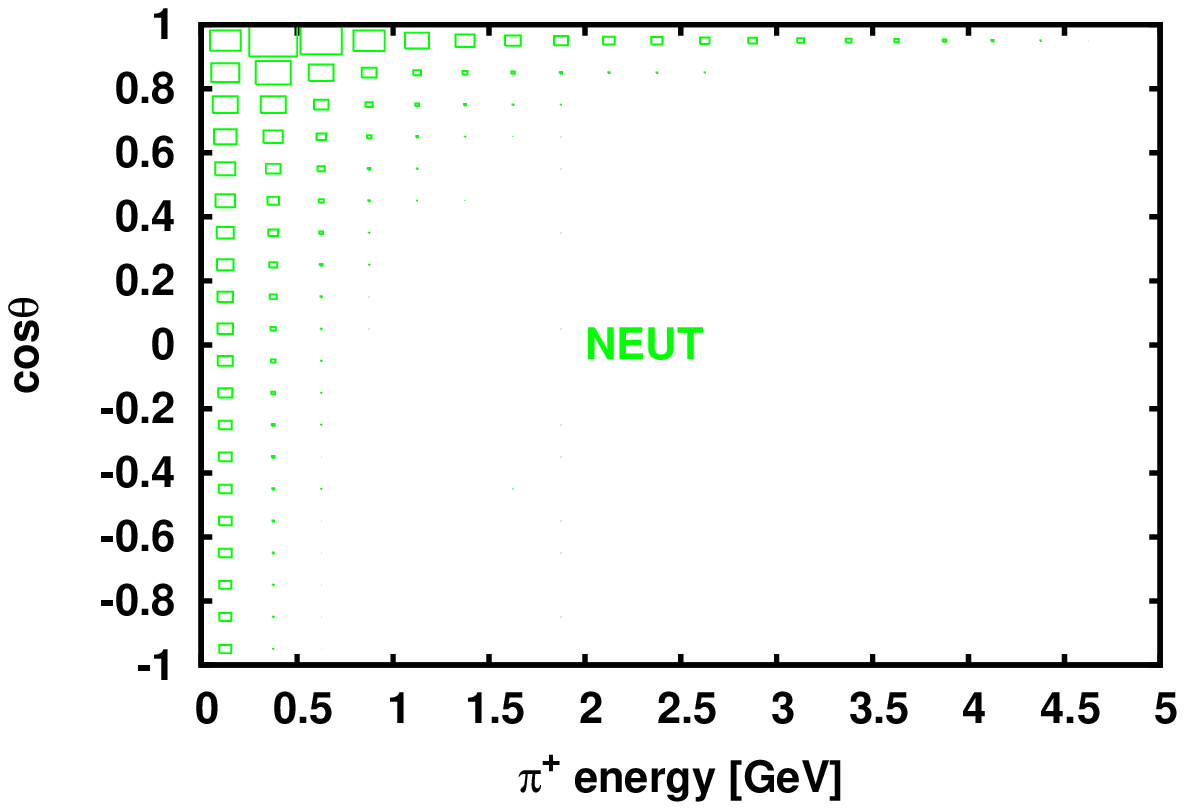}}}
\end{tabular}

\label{fig: H31}
\caption{$\pi^+$ total energy vs cosine of scattering angle distribution in the case of $5$~GeV neutrino-carbon CC interaction. Only event with single $\pi^+$ (and no other pions) are taken into account. The field of the boxes are proportional to the cross section.}

\end{figure}

Finally, $\pi^+$ two-dimensional energy vs cosine of scattering angle (respect to the neutrino beam) distribution in the case of $5$~GeV neutrino CC scattering off carbon, for events with single $\pi^+$ (and no other pions) in the final state is presented in Fig.~\ref{fig: H31}.

The pions cascade models were intensively studied in the last few years in MC generators, as events with pion production in the primary vertex and its absorption during a propagation through nucleus are important background for quasi-elastic interactions. On the plots one can see much better agreement between predictions as for nucleons observables.

GENIE and NuWro results are very similar, but GENIE predicts larger smearing of the distribution. NEUT has a sharp peak around $T_k ~ 0.5$~GeV for forward moving pions and that is why it predicts smaller number of events with more energetic or backward moving pions.

The distribution is sensitive to both: pion production process from primary vertex and pions cascade model. It would be interesting to compare the predictions in the absence of the FSI effects.

\section{Conclusions}

The predictions of four MC neutrino event generators for four observables are investigated. The general conclusion is that the results are not consistent, what should be a subject of more careful investigation to understand the reasons for that. The physics underlying the numerical implementations is the same and the general idea of the construction of the MC codes is similar, so the predictions are expected to agree better. 

Such investigation should start at the point where the interaction happens - in the primary vertex, what would remove the uncertainties related to the final state interactions.

As far as the FSI models testing is concerned, one should start from a few simple observables, like nucleon transparency (a number of nucleons leaving nucleus without any interactions), nucleon multiplicities, energy loss due to FSI effects, or kinematics for the elastic nucleon-nucleon scattering or for the pion absorption process.

Nucleons may carry a useful informations for analysis a neutrino-nucleus scattering, especially in the case of NC interactions, where there is no charged lepton in a final state. Nucleons cascade was not studied as intensively as pions FSI in MC neutrino event generators and from the results presented in this paper it becomes clear, that the closer investigation is needed. 

\begin{theacknowledgments}

The author was partially supported by the grant No. 4585/PB/IFT/12, UMO-2011/M/ST2/02578. I would like to thank to Hugh Gallagher, Yoshinari Hayato, and Sam Zeller for preparing MC predictions for GENIE, NEUT, and NUANCE.

\end{theacknowledgments}

\bibliographystyle{aipproc}
\bibliography{golan_nuint12_ref.bib}

\end{document}